\documentclass[letter,twocolumn]{jpsj2}

\def\runauthor{}

\title{%
Emergent Nodal Excitations due to the Coexistence of Superconductivity and
Antiferromagnetism: Cases with and without Inversion Symmetry
}

\author{%
Satoshi Fujimoto\thanks{E-mail address: fuji@scphys.kyoto-u.ac.jp}
}

\inst{%
Department of Physics, Kyoto University, Kyoto 606-8502, Japan
}

\recdate{\today}

\abst{%
We argue the emergence of nodal excitations due to the coupling with
static antiferromagnetic order 
in fully-gapped superconducting states in 
both cases with and without inversion symmetry.
This line node structure is not accompanied with the sign change of
the superconducting gap, in contrast to usual unconventional Cooper pairs
with higher angular momenta. 
In the case without inversion symmetry, the stability of the
nodal excitations 
crucially depends on the direction of the antiferromagnetic
staggered magnetic moment. 
A possible realization of this phenomenon in CePt$_3$Si is discussed.
}

\kword{ superconductivity, antiferromagnetism, line node, inversion symmetry
}

\begin{document}
\maketitle

The interplay between magnetism and superconductivity has been 
one of central issues in the field of strongly correlated electron systems.
It has been argued that in some heavy fermion systems 
such as CeCoIn$_5$,\cite{ce1} 
CeRhIn$_5$,\cite{crh} 
and UPd$_2$Al$_3$,\cite{upd,upd2} 
antiferromagnetic (AF) critical spin fluctuations may mediate
unconventional Cooper pairs
of which the superconducting (SC) gap has
line nodes.\cite{rev}
Moreover, the coexistence of the AF order and the SC state
is realized in CeRhIn$_5$, UPd$_2$Al$_3$, 
CePt$_3$Si,\cite{bau} CeRhSi$_3$,\cite{kim} 
and CeIrSi$_3$.\cite{onuki}
Among them, CePt$_3$Si, a superconductor without inversion symmetry, 
does not display magnetically critical behaviors, indicating
that the spin fluctuation is almost suppressed, and the static AF order
is well stabilized.\cite{bau} However, this system exhibits some curious
properties described as follows.
The NMR experiments for this system 
indicate the existence of the coherence peak of $1/(T_1T)$,
implying the realization of the fully-gapped state.\cite{yogi}
In contrast, the experimental measurements of thermal transport in 
magnetic fields,\cite{iza}
and the penetration depth,\cite{pene} 
strongly support the line-node structure of the low-energy excitations.
To explain these observations, the present author\cite{fuji2} 
and Hayashi et al.\cite{haya}
pointed out independently that the coherence factor of $1/(T_1T)$
is enhanced in the absence of inversion symmetry, which leads
the prominent coherence peak of $1/(T_1T)$. 
Also, the present author proposed that 
the line node may stem from the coupling with the AF order; i.e.
it is an accidental line node which is not associated with the symmetry of
the Cooper pair,\cite{fuji2} while Hayashi et al. attribute 
the existence of the line node
to the admixture of the spin singlet and triplet pairs.\cite{haya}

In this paper, we further 
pursue the possible realization of nodal excitations in SC states 
{\it due to the coupling with the static antiferromagnetic order}.
We would like to stress that this nodal structure is quite different from
the line-nodes caused by dynamical spin fluctuations as realized in
CeCoIn$_5$ and UPd$_2$Al$_3$ in the point that
the former is not accompanied with the sign change of
the SC gap at the line-nodes.

To explain the basic properties of this phenomenon,
we, first, consider the case of the $s$-wave pairing 
with inversion symmetry coexisting with AF order.
Although systems corresponding to this situation have not been 
realized in any real compounds so far,
the consideration of this case is useful for 
a simple theoretical description of
the mechanism for the emergent line-node due to the AF order.
We start from the following mean field Hamiltonian,
\begin{eqnarray}
H&=&\sum_{k,\sigma}\varepsilon_k c^{\dagger}_{k\sigma}c_{k\sigma}
+\sum_{k,\sigma.\sigma'}[\Delta^{(s)} 
(i\sigma_y)_{\sigma\sigma'}c_{k\sigma}c_{-k\sigma'}+h.c.]
\nonumber \\
&+&
\sum_{k,\sigma,\sigma'}[\mbox{\boldmath $m$}_Q\cdot
\mbox{\boldmath $\sigma$}_{\sigma\sigma'}
c^{\dagger}_{k+Q \sigma}c_{k \sigma'}+h.c.]. \label{ham1}
\end{eqnarray}
Here, $c_{k\sigma}$ ($c^{\dagger}_{k\sigma}$) is the annihilation (creation)
operator of an electron with momentum $k$, and spin $\sigma$.
$\Delta^{(s)}$ is the $s$-wave SC gap 
independent of the momentum $k$, 
and  $\mbox{\boldmath $m$}_Q$ is the staggered
magnetic moment.
We do not specify the origin of the AF order in the following argument.
It may stem from localized electrons which are not explicitly included in 
(\ref{ham1}).  
The single-particle excitation energy of (\ref{ham1}) is readily obtained as,
\begin{eqnarray}
&&E_{k\pm}=\biggl[\frac{\varepsilon^2_k+\varepsilon^2_{k+Q}}{2}+\Delta^{(s)2}_{k}
+|\mbox{\boldmath $m$}_Q|^2 \nonumber \\
&&\pm\biggl[\frac{(\varepsilon_k^2-\varepsilon_{k+Q}^2)^2}{4}
+|\mbox{\boldmath $m$}_Q|^2(\varepsilon_k+\varepsilon_{k+Q})^2 
+4|\mbox{\boldmath $m$}_Q|^2\Delta^{(s)2}_{k}\biggr]^{\frac{1}{2}}
\biggr]^{\frac{1}{2}}.\label{eneinv}
\end{eqnarray}
Apparently, in the case of $|\mbox{\boldmath $m$}_Q|\gg\Delta^{(s)}$, 
which corresponds to the situation realized in CePt$_3$Si,
excitations with the energy $E_{k+}$ 
are not important for low-energy properties.
Thus we ignore their contributions in the following. 
The SC gap $\Delta^{(s)}$ should be determined by solving 
the gap equation self-consistently.
Assuming the BCS-type attractive interaction between electron pairs, 
$-V\sum_{k,k'} c^{\dagger}_{k\uparrow}c^{\dagger}_{-k\downarrow} 
c_{-k'\downarrow}c_{k'\uparrow}$, we obtain 
the gap equation,
\begin{eqnarray}
&&1=V\sum_k\frac{\tanh\left(\frac{E_{k}}{2T}\right)}{2E_k}\biggl(1 
-\frac{\frac{\varepsilon_k^2-\varepsilon_{k+Q}^2}{2}
+2|\mbox{\boldmath $m$}_Q|^2}
{E_{k+}^2-E_{k-}^2}\biggr). \label{gap}
\end{eqnarray}
In the vicinity of the Fermi surface which yields 
the most dominant contributions
to the gap equation,
the factor in the parenthesis 
of Eq.(\ref{gap}) is non-negative, and thus
the right-hand side of the gap equation is always positive for $V>0$.
This means that for any given values of $\Delta^{(s)}$ 
and $|\mbox{\boldmath $m$}_Q|$,
there exists a positive value of $V>0$ which satisfies
the gap equation (\ref{gap}).
Therefore, in the following, 
we treat $\Delta^{(s)}$ and $|\mbox{\boldmath $m$}_Q|$ as independent
parameters, without solving the gap equation explicitly.
To simplify the calculation below, we assume that 
$\varepsilon_k=k^2/(2m)-E_F$ with $m$ the electron mass and 
$E_F$ the Fermi energy, and that the ordering wave vector
of the antiferromagnetism is $\mbox{\boldmath $Q$}=(0,0,\pm\pi)$.
For sufficiently large magnitude of $|\mbox{\boldmath $m$}_Q|$, 
the AF order strongly affects the electronic structure 
in the vicinity of the magnetic Brillouin Zone at $k_z=\pm\pi/2$,
and eventually, destroys
the single-particle excitation gap due to the SC order.
We demonstrate this phenomenon numerically for a particular set of parameters.
The results are shown in Fig.1.
The height of the hills indicates the magnitude of the excitation gap,
which vanishes
around the magnetic Brillouin Zone
for $|\mbox{\boldmath $m$}_Q|\gg \Delta^{(s)}$,
indicating the line-node structure.
We also calculate the local density of states 
$D(\varepsilon)=-\frac{1}{\pi}\sum_k{\rm Im}G_k^R(\varepsilon)$,
where $G^R_k(\varepsilon)$ is the retarded normal Green function of the
Hamiltonian (\ref{ham1}). 
We show $D(\varepsilon)$ plotted as a function of the energy $\varepsilon$
in Fig.2.
For sufficiently small energies $\varepsilon \ll \Delta^{(s)}$,
the density of states is linear in $\varepsilon$, exhibiting
the line-node-like behavior.
The existence of the nodal excitations in the $s$-wave pairing state
is understood as follows.
In the vicinity of the magnetic Brillouin Zone,
we apply an approximation $\varepsilon_{k+Q}\approx -\varepsilon_k$.
Then, the single-electron excitation energy (\ref{eneinv}) is recast into
$
E_k=\sqrt{\varepsilon_k^2+(\Delta^{(s)}-|\mbox{\boldmath $m$}_Q|)^2}.
$
Thus, when $\Delta^{(s)}=|\mbox{\boldmath $m$}_Q|$, 
the single-electron excitation
energy is gapless.
More precisely, since the approximation 
$\varepsilon_{k+Q}\approx -\varepsilon_k$
becomes worse for $\mbox{\boldmath $k$}$ away from the magnetic Brilloin Zone,
it is required to tune 
the values of $|\mbox{\boldmath $m$}_Q|$ much larger than $\Delta^{(s)}$
for the realization of the nodal excitations. From this consideration, 
it is clear that the SC gap itself does not change 
the sign at this line node.
Moreover, this node-like structure is not a true line node, but 
a minimum of the excitation gap. 
Nevertheless, the overall behavior of the density of states is 
quite similar to
that of usual SC states with line nodes, and also
any thermodynamic quantities at low temperatures behaves as if
the true line node exists.

\begin{figure}
\begin{center}
\includegraphics[width=8cm]{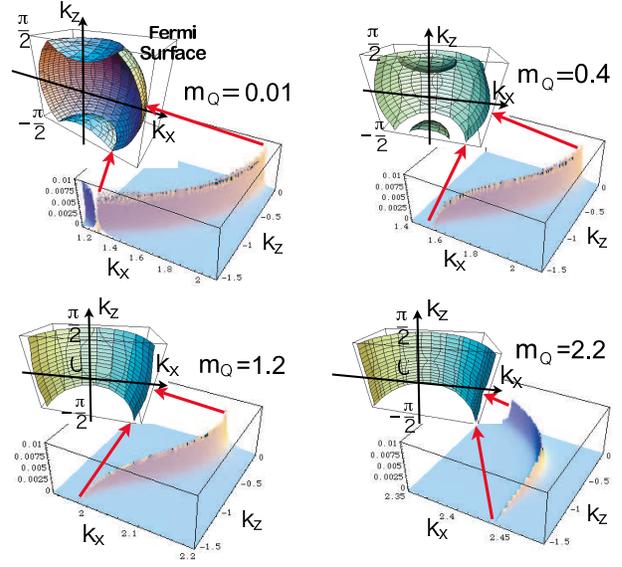}
\end{center}
\caption{(Color online) 
The single-particle excitation gap plotted as a function of $k_x$ 
and $k_z$ for $|\mbox{\boldmath $m$}_Q|=0.01, 0.4, 1.2, 2.2$. 
Here, we put the electron mass $m=1.0$, the Fermi energy $E_F=1.0$, 
and the SC gap 
$\Delta^{(s)}=0.01$.
Figures of a portion of the Fermi surface folded 
by the coupling with the AF order
are also shown.}
\label{f1}
\end{figure}

\begin{figure}
\begin{center}
\includegraphics[width=5cm]{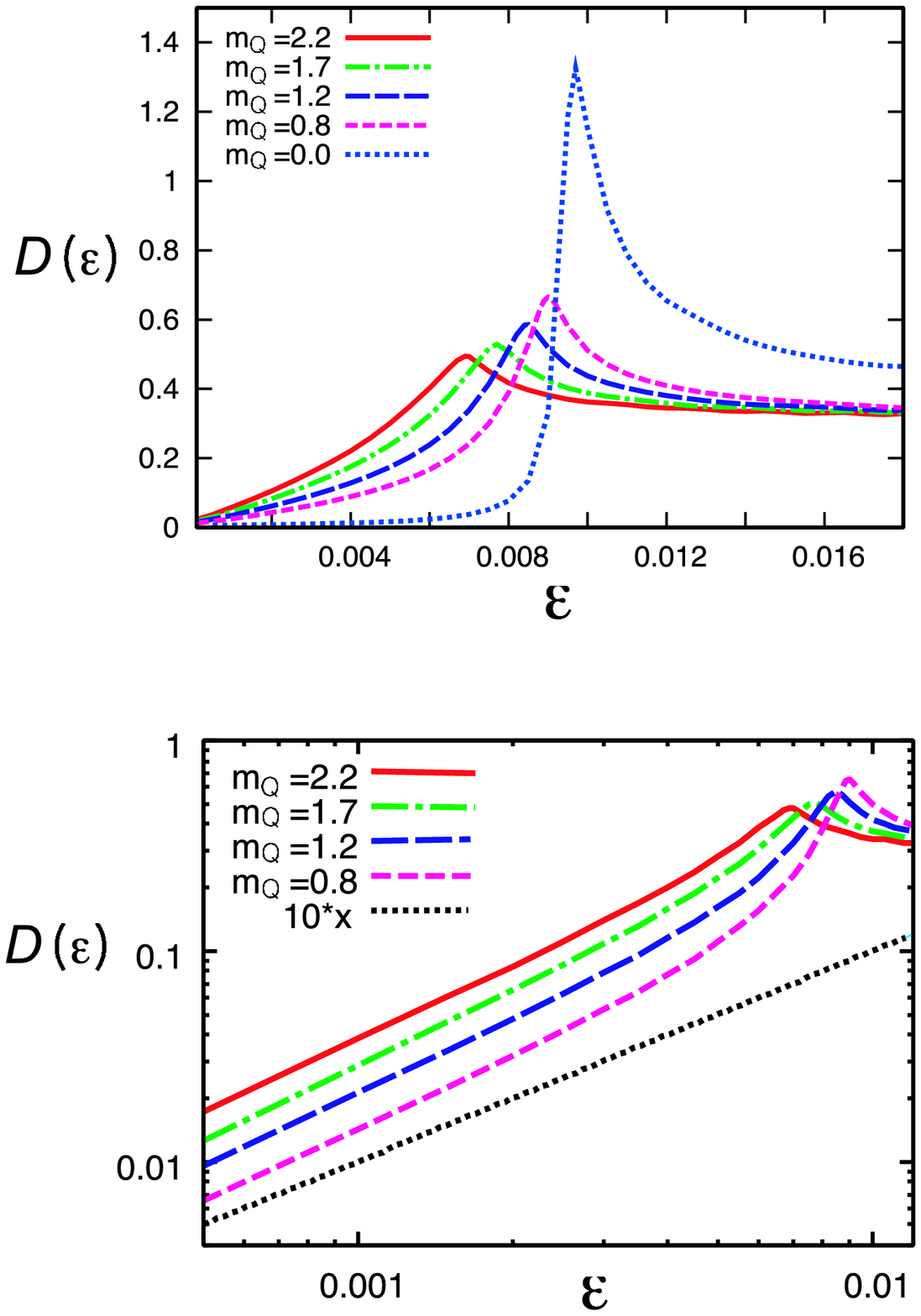}
\end{center}
\caption{(Color online) The local density of states for the model (\ref{ham1}) 
plotted against 
the excitation energy $\varepsilon$. The upper panel is
in the linear scale. The lower panel is a log-log plot.}
\label{f2}
\end{figure}

\begin{figure}
\begin{center}
\includegraphics[width=8cm]{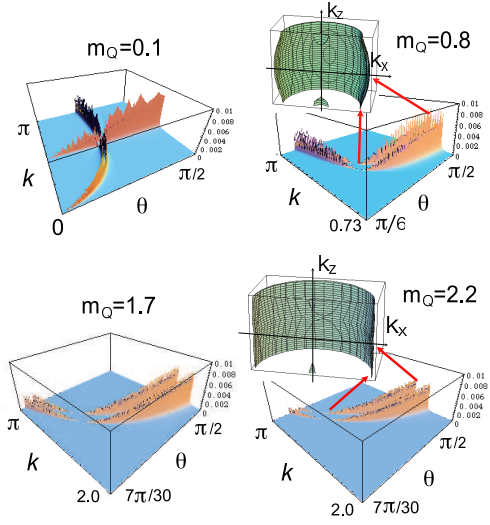}
\end{center}
\caption{(Color online) 
The single-particle excitation gap plotted as a function of $k$ 
and $\theta$ for $|\mbox{\boldmath $m$}_Q|=0.01, 0.8, 1.7, 2.2$ in the case 
of the $p$-wave pairing state without inversion symmetry and
$\mbox{\boldmath $m$}_Q\parallel x$. 
Here, we put 
$m=1.0$, 
$E_F=1.0$, 
$\alpha=0.07$,
and 
$\Delta^{(t)}=0.01$.
Figures of a portion of the Fermi surfaces 
are also shown.}
\label{f3}
\end{figure}

\begin{figure}
\begin{center}
\includegraphics[width=5cm]{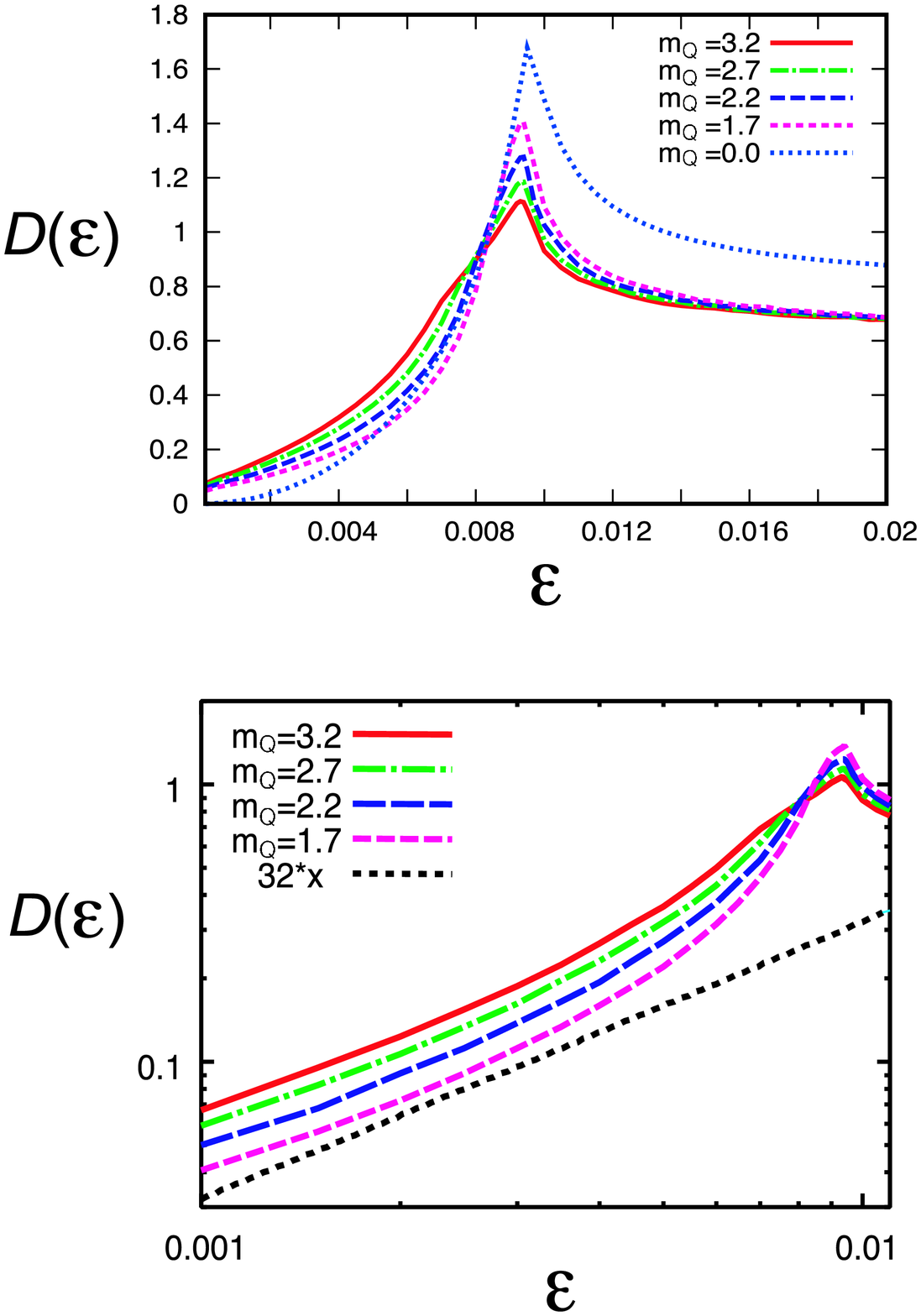}
\end{center}
\caption{(Color online) The local density of states 
for the model (\ref{ham2}) plotted against 
the excitation energy $\varepsilon$. The upper panel is
in the linear scale. The lower panel is a log-log plot.}
\label{f4}
\end{figure}

\begin{figure}
\begin{center}
\includegraphics[width=7cm]{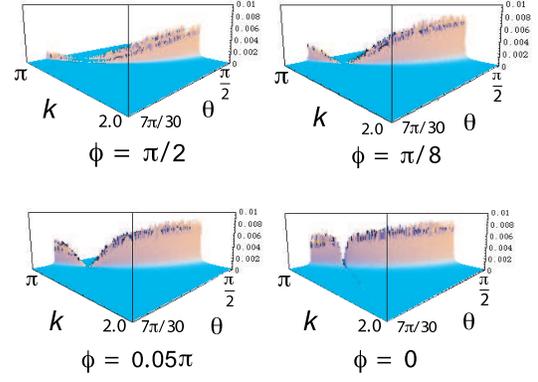}
\end{center}
\caption{(Color online) 
The excitation gap plotted as a function of $k$ and $\theta$
for $\phi=\pi/2,\pi/8,0.05\pi,0$ for the model (\ref{ham2}).
The line node structure disappears around $\phi\sim 0$.}
\label{f5}
\end{figure}

\begin{figure}
\begin{center}
\includegraphics[width=2.7cm]{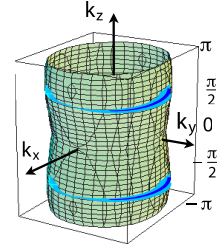}
\end{center}
\caption{(Color online) 
A schematic figure of the line node structure on the Fermi surface.
The depth of the blue color in the vicinity of the magnetic Brillouin Zone
$k_z\pm\pi/2$ indicates the depth of the node of the excitation gap.}
\label{f6}
\end{figure}

\begin{figure}
\begin{center}
\includegraphics[width=7cm]{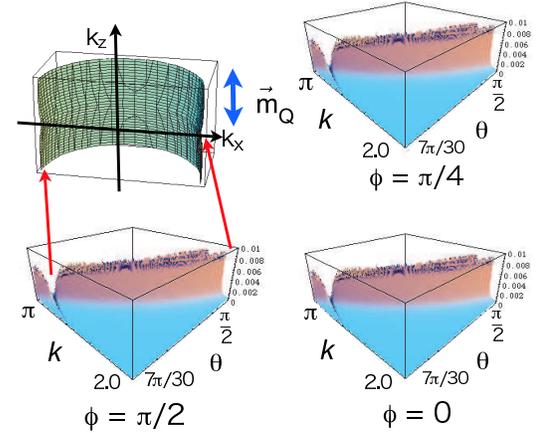}
\end{center}
\caption{(Color online) The excitation gap for the model (\ref{ham2}) for 
$|\mbox{\boldmath $m$}_Q|=2.7$
in the case of $\mbox{\boldmath $m$}_Q\parallel z$-axis
that
the AF staggered moment is parallel to the $z$-axis.}
\label{f7}
\end{figure}

The above mechanism for the emergence of the line nodes is also applicable to
systems without inversion symmetry.
In this case, parity violation allows the admixture of 
the spin-singlet and triplet states.\cite{ede,gr}
The ratio of the minor component of the SC gap
induced by the spin-orbit (SO) interaction
to the major component is of order $\delta E_{SO}/E_F$ where $\delta E_{SO}$
is the magnitude of the spin-orbit splitting of the energy band.
In any superconductors without inversion symmetry discovered so far, this ratio
is less than $\sim 0.1$.\cite{bau,kim,onuki} 
Thus, in the following, we neglect the minor component of the SC gap.
As explained above, for CePt$_3$Si,
experimental observations seems to support the realization of 
the $s+p$-wave state.\cite{yogi,fuji2,fri}
Since 
the on-site Coulomb repulsion 
should be considerably large in heavy fermion systems, 
it is appropriate to assume that 
the major component 
of the SC gap has the $p$-wave symmetry, and ignore 
the minor $s$-wave component. 
We approximate the inversion-symmetry-broken spin-orbit interaction by
the Rashba-type interaction.
Then the model Hamiltonian is given by,
\begin{eqnarray}
H&=&\sum_{k,\sigma}\varepsilon_k c^{\dagger}_{k\sigma}c_{k\sigma} 
+\alpha\sum_{k,\sigma,\sigma'}(\mbox{\boldmath $k$}\times\mbox{\boldmath $n$})
\cdot\mbox{\boldmath $\sigma$}_{\sigma\sigma'}
\nonumber \\
&+&\sum_{k,\sigma.\sigma'}[\Delta^{(p)} (\mbox{\boldmath $k$}
\times\mbox{\boldmath $n$})
\cdot(\mbox{\boldmath $\sigma$}
i\sigma_y)_{\sigma\sigma'}c_{k\sigma}c_{-k\sigma'}+h.c.] \nonumber \\
&+&
\sum_{k,\sigma,\sigma'}[\mbox{\boldmath $m$}_Q\cdot
\mbox{\boldmath $\sigma$}_{\sigma\sigma'}
c^{\dagger}_{k+Q \sigma}c_{k \sigma'}+h.c.]. \label{ham2}
\end{eqnarray}
Here, the second term of the right-hand side of 
(\ref{ham2}) is the Rashba spin-orbit interaction, 
$\Delta^{(p)}$ is the amplitude of the SC gap.
Note that in the above model 
the $\mbox{\boldmath $d$}$-vector of the triplet pairing 
is determined by the Rashba spin-orbit interaction. 
Also, we assume that the magnetic ordering wave vector 
is $\mbox{\boldmath $Q$}=(0,0,\pm\pi)$, and that 
the AF staggered moment is directed along the $x$-axis,i.e.
$\mbox{\boldmath $m$}_Q=(|\mbox{\boldmath $m$}_Q|,0,0)$,
in accordance with the results of 
the neutron scattering measurements for CePt$_3$Si, 
which suggest the existence of the in-plane magnetic moment.\cite{neu}
As will be shown below, the relative angle between the direction of
the staggered moment and that of the inversion-symmetry-breaking 
potential gradient
($\mbox{\boldmath $n$}$ in Eq.(\ref{ham2})) is crucially important
for the emergence of the nodal excitations
in the spin triplet dominated case.  
To simplify the following analysis, 
we assume $\varepsilon_k=k^2/(2m)-E_F$, again.
Although the actual band structure of CePt$_3$Si is much more complicated,
the essential feature of the emergent line node due to the AF order
is not affected by this simplification.
Since the single-electron energy of (\ref{ham2}) 
can not be obtained analytically,
we diagonalize the Hamiltonian (\ref{ham2}) numerically 
for a particular set of parameters.
The calculated results of the excitation energy gap is plotted in Fig.3, 
in which
we use the spherical coordinate in the momentum space 
$\mbox{\boldmath $k$}=(k\cos\phi\sin\theta,k\sin\phi\sin\theta,k\cos\theta)$.
It is seen that the Fermi surfaces are splitted into 
two pieces by the SO interaction.
In the vicinity of the magnetic Brillouin Zone at $k_z=\pm\pi/2$,
the excitation gap is collapsed on both the splitted Fermi surfaces.
The local density of states for this model is plotted as a function of
the excitation energy in Fig.4.
When the magnitude of $|\mbox{\boldmath $m$}_Q|$ is much larger than
the SC gap, 
the density of states obeys
$D(\varepsilon) \propto \varepsilon$ for $\varepsilon\ll \Delta^{(t)}$, 
indicating the line-node-like structure
of the low-energy excitations.
However, for higher energies $\varepsilon\sim\Delta^{(t)}$, 
the energy dependence of the density of states differs from
the line-node behavior, and exhibit a prominent peak structure around 
$\varepsilon\sim\Delta^{(t)}$, as in the case of fully-gapped states.
This peak structure may be important for
the realization of the coherence peak of $1/(T_1T)$ observed for 
CePt$_3$Si.\cite{yogi}
Since the SC gap does not change the sign at this nodal line,
the coherence factor gives substantial contributions to $1/(T_1T)$,
resulting in the enhancement of the coherence peak, in agreement with
the experimental observations.\cite{yogi}
We would like to note that for sufficiently large $|\mbox{\boldmath $m$}_Q|$,
the point nodes of the $p$-wave gap at $k_x=0$, $k_y=0$ disappear,
because of the deformation of the Fermi surface as depicted in Fig.3.
This fact is also favorable for the enhancement of the coherence peak.

Another interesting feature of this nodal structure is that
it possesses the $C_{2v}$ symmetry 
in the momentum space, 
because of the existence of $\mbox{\boldmath $m$}_Q$ 
aligned to the $x$-axis and
the inversion-symmetry-breaking SO interaction.
This property is clearly seen in Fig.5, in which the variation of the nodal
structure 
as a function of the azimuthal angle $\phi$ is described.
We depict the line-node structure 
with the $C_{2v}$ symmetry on the Fermi surface 
schematically in Fig.6.
The experimental detection of this two-fold symmetry
in CePt$_3$Si is an important test
for the present theory.
Moreover, in contrast to the case of the spin singlet state 
with inversion symmetry considered before,
the emergence of the nodal excitations 
in the triplet dominated state without inversion symmetry
crucially depends on
the direction of the staggered magnetic moment $\mbox{\boldmath $m$}_Q$.
To see this, we display the excitation gap calculated by assuming 
$\mbox{\boldmath $m$}_Q=(0,0,|\mbox{\boldmath $m$}_Q|)$ 
with $|\mbox{\boldmath $m$}_Q|=2.7$ in Fig.7.
The line-node structure does not appear in this case.
This observation indicates that 
in the triplet pairing dominated case,
the emergence of the nodal excitations, or, conversely, the suppression
of the SC gap due to the coupling with the AF order
occurs only when the magnetic moment orthogonal to
the spin of the triplet Cooper pair exists, which disturbs
the formation of the triplet pair.

Finally, we make a brief comment on the universal conductivity
of the nodal excitations perturbed by random impurity potentials.
It is known that low-energy excitations from line nodes gives
a universal value of the thermal conductivity due to impurity scattering 
in the zero temperature limit.\cite{lee}
Actually, the universal thermal conductivity 
is observed for CePt$_3$Si.\cite{iza}
For the emergent line nodes considered here, the universal conductivity
also appears provided that $|\mbox{\boldmath $m$}_Q|$ is much larger than
the SC gap, as verified by the straightforward calculation 
for the models (\ref{ham1}) and (\ref{ham2}).

In summary, we have presented a mechanism for the emergence 
of line-node structures in fully-gapped SC states due to
the coupling with static AF order.
The possible realization of this phenomenon
in CePt$_3$Si has been proposed to reconcile
the existence of the coherence peak of $1/(T_1T)$ with 
the experimental observations of nodal excitations.

\acknowledgment{}

The author is grateful to K. Yamada, 
Y. Matsuda, H. Mukuda, M. Yogi, and H. Ikeda for invaluable discussions.
This work was partly supported by a Grant-in-Aid from the Ministry
of Education, Science, Sports and Culture, Japan.

\end{document}